\documentclass[12pt]{article}
\input amssym.def
\input amssym.tex

\begin{document}

\title{PAIR CORRELATIONS OF QUANTUM CHAOTIC MAPS FROM SUPERSYMMETRY}

\author{M R Zirnbauer%
\footnote{Inst. f. Theor. Physik, Universit\"{a}t K\"{o}ln, Germany; 
email: \texttt{zirn@thp.Uni-Koeln.DE}}}

\date{November 15, 1997}
\maketitle

\begin{abstract} 
  A conjecture due to Bohigas, Giannoni and Schmit (BGS), stating that
  the energy level correlations of quantum chaotic systems generically
  obey the laws of random matrix theory, is given a precise
  formulation for quantized symplectic maps.  No statement is made
  about any individual quantum map.  Rather, a few--parameter ensemble
  of maps is considered, such that the deterministic map is composed
  with a diffusion operator on average.  The ensemble is a ``quantum''
  one, which is to say that the diffusion operator contracts to the
  identity in the classical limit.  It is argued that the BGS
  conjecture is true on average over such an ensemble, provided that
  the classical map is mixing.  The method used is closely related to
  the supersymmetric formalism of Andreev et al for chaotic
  Hamiltonian systems.
\end{abstract}

\section{Introduction}

One of the goals of semiclassical analysis is to characterize the
energy level correlations of quantum mechanical systems in the limit
$\hbar \to 0$.  To gain an understanding of these correlations,
physicists have traditionally used semiclassical trace formulas,
relating the density of energy levels to the periodic orbits of the
classical dynamics (see e.g. the contributions by Bogomolny, Keating,
and Smilansky to this volume).  In brief, the qualitative picture that
has emerged is that integrable systems have Poisson statistics,
whereas for a generic chaotic system one expects random matrix (or
Wigner--Dyson) statistics.  This picture is supported by a large body
of numerical evidence, and applies equally to Hamiltonian systems and
symplectic maps.  Bohigas, Giannoni, and Schmit (BGS) \cite{bgs} are
credited for having put forth the random matrix conjecture for quantum
chaotic systems.  The principal development toward an analytical
theory is due to Berry \cite{berry} who augmented Gutzwiller's trace
formula with sum rule arguments to determine the form factor for times
much shorter and much longer than the Heisenberg time.  Berry's work
has recently been refined by Bogomolny and Keating \cite{bk}.

In these lectures, I will describe a supersymmetric formalism that
offers an alternative to the trace formula approach and promises
rigorous results for chaotic symplectic maps.  The formalism bears
much similarity to that of Andreev et al \cite{a3s} for chaotic
Hamiltonian systems, reviewed at this workshop by Ben Simons.  Only
the simplest case of maps without unitary or anti--unitary symmetries
will be treated here.  In particular, time reversal invariance will be
assumed to be broken.  The starting point of our theory is an
expression for the two-level correlation function which involves two
ratios of quantum spectral determinants (Sect.~\ref{sec:conjecture}).
The determinants are written as traces over a Fock space of fermions,
and the inverse determinants as traces over a Fock space of bosons
(Sect.~\ref{sec:formula_1}).  By the introduction of generalized
coherent states, the two-level correlation function can then be
expressed as a Berezin integral over the Riemannian symmetric
superspace of type $A{\rm III}|A{\rm III}$
(Sect.~\ref{sec:formula_2}).  A quick review of the basic mathematics
underlying the Berezin integral is given in Sect.~\ref{sec:basics}.

The BGS conjecture is not a precise statement, as the meaning of the
word ``generic'' is left undefined.  (The qualification ``generic''
serves to exclude some prominent counterexamples, such as arithmetic
billiards and the cat maps, which are paradigms of classical chaos but
fail to obey random matrix statistics when quantized in the canonical
way.)  It turns out that in our formalism it is impossible to prove
the conjecture in the sense originally intended, i.e. as a statement
about an individual quantum system.  The technical obstacle that
prevents us from making progress for individual systems is the
nonexistence of the semiclassical limit (Sect.~\ref{sec:semiclassics})
for the Berezin integral representation of the two-level correlation
function, or any other correlation function.  To enforce such a limit
one needs to impose some sort of coarse graining, or ultraviolet
regulator, on the Berezin integral.  The way to implement
regularization is to compose the map with a stochastic Hamiltonian
flow and average (Sect.~\ref{sec:regularization}).

From experience with the supersymmetry formalism applied to disordered
electron systems \cite{efetov} it is known that the so--called
zero mode approximation to the Berezin integral gives the random
matrix answer.  The strategy therefore is to establish sufficient
conditions for this approximation, which is of the saddle point type,
to be justified.  The key issue now becomes the stability of the
manifold of saddle points.  Without ensemble averaging, the Hessian of
the saddle point manifold is not positive, and the saddle point
approximation is false.  Averaging over the stochastic Hamiltonian
flow with ``time'' parameter $\epsilon$ stabilizes the saddle point
manifold and leads to a good asymptotic expansion
(Sect.~\ref{sec:asymptotics}).  If the classical map is mixing we
argue from power counting that stability can achieved with a time
parameter $\epsilon$ that vanishes with $\hbar$ as $\hbar^\alpha$ where
$0 < \alpha < 1$.  Thus we propose that the BGS conjecture holds on
average over an ensemble of quantum chaotic maps {\it all of which
have the same classical limit}.

My original intention was to finish with a discussion of the finite
$\hbar$ corrections to the universal random matrix result.  These are
a very timely issue, as the ballistic nonlinear $\sigma$ model
calculation of Andreev et al disagrees with the result of Bogomolny
and Keating derived from the diagonal approximation.  (The difference
lies in the weights given to the contributions from periodic orbits
with repetition number $r \ge 2$.)  There exist strong indications
that the result of Bogomolny and Keating is the correct one.  I
believe that the resolution of the discrepancy will teach us
something profound about the ballistic nonlinear $\sigma$ model, but
this definitely exceeds the scope of the present lectures and must be
left for a future publication.

\section{Preliminaries}

We begin with some basic definitions.

Let $M$, called the phase space, be a $2d$--dimensional compact
manifold with a symplectic structure $\omega$, i.e. an antisymmetric
second rank tensor that is expressed in local coordinates $q_k,\;p_k$
by $\omega =\sum_{k=1}^d{\rm d}p_k\wedge {\rm d}q_k$. Via $\omega$,
every function $f:M\to {\Bbb R}$ is associated with a Hamiltonian
vector field $\Xi_f =\sum_{k=1}^d \left(\frac{\partial f}{\partial
    p_k} \frac{\partial}{\partial q_k}-\frac{\partial f}{\partial q_k}
  \frac{\partial}{\partial p_k}\right)$. The volume element, or
Liouville measure, on $M$ is denoted by $dx:=\omega^{\wedge d}$. A map
$\chi : M \to M$ is called symplectic (or area preserving for $d=1$)
if it preserves $\omega$ and, consequently, the Liouville measure.
The map $\chi$ acts on functions $f:M\to {\Bbb R}$ by $(\chi^{\ast}
f)(x)=f(\chi (x))$. The dual of its inverse, ${\chi^{-1}}^{\ast}$, is
called the Frobenius--Perron operator. A little computation,
          \begin{eqnarray}
          &&({\chi^{-1}}^{\ast}f)(x)=f(\chi^{-1}(x))
          \nonumber \\
          &=&\int_M dy \,
          \delta (\chi^{-1}(x)-y)f(y)=\int_M dy \,
          \delta (x-\chi (y))f(y) \,,
          \nonumber
          \end{eqnarray}
shows that the Frobenius--Perron operator has the integral kernel
          \[
          ({\chi^{-1}}^{\ast})(x,y)=\delta (x-\chi (y))\,,
          \]
which identifies ${\chi^{-1}}^{\ast}$ as the propagator (for one time
step) of the classical dynamics obtained by iterating $\chi$.

Quantization turns every smooth function $f:M\to {\Bbb R}$ into a
self--adjoint operator ${\cal Q}_N(f)=F_N$ on a Hilbert space ${\cal
  H}_N$ of finite dimension $N = (2\pi\hbar )^{-d}\int_M dx$.  The
quantization of a smooth phase space function is called a
pseudodifferential operator.  The quantum analog of the classical map
$\chi$ is a unitary operator $U_N:{\cal H}_N\to {\cal H}_N$, whose
eigenvalues we denote by $e^{i\theta_n}\; (n=1,\ldots ,N)$.  The
$\theta_n$'s are sometimes called quasienergies.  Let ${\cal D}_N =
{\cal H}_N\otimes {\cal H}_N^{\ast}\simeq {\rm End}({\cal H}_N)$ be
the space of linear operators on ${\cal H}_N$. The unitary operator
$U_N$ acts on ${\cal D}_N$ by conjugation. More precisely, the quantum
time evolution (by one time step) of an element $F_N \in {\cal D}_N$
is given by
          \[
          {\rm Ad}(U_N)^{-1}F_N=U_N^{-1}F_NU_N\;.
          \]
(The notation ${\rm Ad}(U)$ for  conjugation by $U$ is taken from
Lie algebra theory.) In the classical limit $\hbar\to 0$, 
${\rm Ad}(U_N)^{-1}$ acting on a pseudodifferential operator
$F_N={\cal Q}_N(f)$ approaches the Frobenius--Perron operator:
          \[
          {\rm Ad}(U_N)^{-1}F_N\;\stackrel{\hbar\to 0}{\longrightarrow}\;
          {\cal Q}_N({\chi^{-1}}^{\ast}f)\;.
          \]

Examples illustrating these general definitions are furnished by the
cat maps \cite{arnold}.  The phase space in this case is the
two--torus ${\rm T}^2$, coordinatized by a pair of canonically
conjugate position and momentum functions $q$ and $p$, which are
defined modulo 1.  For a concrete example, consider the cat map $\chi
: {\rm T}^2\to {\rm T}^2$ that acts on $q$ and $p$ as
          \begin{eqnarray*}
          q\circ\chi & = & {\phantom{1}}q + 2p\quad ({\rm mod}\; 1)\,,\cr
          p\circ\chi & = & 2q + 5p\quad ({\rm mod}\; 1)\,,
          \end{eqnarray*}
which entails a Frobenius--Perron operator expressed by
          \[
          {\chi^{-1}}^{\ast}=\exp{(-2q\partial /\partial p)}\, 
          \exp{(-2p\partial /\partial q)}
          \]
on differentiable functions. [Note that $\exp{(-\partial /\partial
  q)}= \exp{(-\partial /\partial p)}=1$, so that $\exp{(-2p\partial
  /\partial q)}$ and $\exp{(-2q\partial/\partial p)}$ are globally
well--defined, although $p\partial/\partial q$ and $q\partial/
\partial p$ are not.]  The locally defined vector fields $\Xi_f =
2q\partial/\partial p$ and $\Xi_g = 2p\partial/\partial q$ derive from
locally defined functions $f = -q^2$ and $g = p^2$, and quantization
of $\chi$ yields a unitary operator
          \[
          U_N = 
          \exp{\left(  iP^2_N / \hbar \right)}\,
          \exp{\left( -iQ^2_N / \hbar \right)}\,,
          \]
where $\hbar^{-1} = 2\pi N\int_0^1 {\rm d}p\int_0^1 {\rm d}q
= 2\pi N$. In a basis where $\exp{(2\pi iP_N)}$ is diagonal with the
eigenvalues being the $N^{\rm th}$ roots of unity $\exp{(2\pi in/N)}\;
(n=1,2,\ldots,N)$, $U_N$ has the following matrix
          \[
          (U_N)_{nn'}=\frac{1}{\sqrt{N}}\exp{\frac{2\pi i}{N}\left( 
          n^2 + (n-n')^2 \right)}\;.
          \]

\section{BGS conjecture formulated precisely}
\label{sec:conjecture}

There exist many statistical measures of level correlations. The one
we will focus on here is the so--called pair correlation function,
defined on a test function $f:{\rm U}(1)\to {\Bbb C}$ by
          \[
          C(f)=\frac{2\pi}{N^2}\sum\limits_{m,n=1}^N f\left(
          e^{i(\theta_m - \theta_{n})}\right) -\int\limits_0^{2\pi}
          f(e^{i\theta}){\rm d}\theta\;.
          \]
If we take the liberty of choosing for $f$ the Dirac 
$\delta$--distribution centered at $e^{i\varphi},\; \delta_{\varphi}
(e^{i\theta})=(2\pi)^{-1}\sum_{l\in {\Bbb Z}} e^{il(\theta -\varphi
  )}$, we get the two--level correlation function
          \[
          R_2(\theta )=C(\delta_{\theta})=\frac{2}{N^2}
          \sum\limits_{l=1}^{\infty}\cos{(l\theta )}
          \left| {\rm Tr}U^l\right|^2\;.
          \]
For our purposes, it is technically convenient to consider in addition
to $R_2$ the following correlator of determinants:
          \[
          \Omega (\gamma_{+0},\gamma_{+1};\gamma_{-0},\gamma_{-1})
          =\int\limits_0^{2\pi}\frac{{\rm d}\theta}{2\pi} \;
          \frac{{\rm Det}(1-\gamma_{+1}e^{i\theta}U_N)
          {\rm Det}(1-\gamma_{-1}e^{-i\theta}U_N^{-1})}{{\rm Det}
          (1-\gamma_{+0}e^{i\theta}U_N)
          {\rm Det}(1-\gamma_{-0}e^{-i\theta}U_N^{-1})}\,,
          \]
where $\gamma_{\pm 0},\gamma_{\pm 1}$ are complex parameters. Note that the
two--level correlation function can be extracted from $\Omega$
by taking two derivatives:
          \[
          R_2(\theta )=-\frac{1}{2N^2}\frac{\partial^2}{\partial\varphi
          \partial\varphi'} {\rm Re} \; \Omega\left( 
          e^{i\theta +i\varphi},e^{i\theta -i\varphi};e^{i\varphi '},
          e^{-i\varphi '}\right)\Big|_{\varphi =\varphi '=0}\;.
          \]
In Sects.~\ref{sec:formula_1}--\ref{sec:formula_2} I will sketch
the derivation of a formula for $\Omega$ that underlies the
supersymmetric method for quantized maps, and I will go over some of
the basic mathematics needed. Before doing so, it is proper to
motivate the reader by giving an outline of the results that can be
obtained.

The first and very important message is that our method {\it gives no
answer at all} for the pair correlation function, or any other
correlation function, of an {\it individual} quantized map. (This is
to be contrasted with what has been claimed by Andreev et al.) For
reasons that will be explained later on, we can make a statement only
about the {\it expected value} of the correlations for an {\it
ensemble} of quantized maps over a well--chosen probability space.

Let me therefore digress and describe this ensemble.  We pick a finite
number $s$ of Hamiltonian functions $x_k : M\to {\Bbb R}$, with
associated Hamiltonian vector fields $\Xi_k$, and Hamiltonian
operators $X_{k,N} : {\cal H}_N\to {\cal H}_N$.  The latter we multiply
by uncorrelated Gaussian random variables $\xi_k$ with zero mean and
$\hbar$--dependent variances $\left<\xi_k^2\right> = 2\epsilon(\hbar
)$, and we consider instead of a single $U_N$ an $s$--parameter family
of unitary operators
           \[
           U_N(\xi ):=\exp{\left( i\sum_{k=1}^s \xi_k X_{k,N}
            / \hbar \right)} U_N\;.
           \]
The choice of Hamiltonians $x_k$ is constrained by the requirement
that the weighted sum of squares $\sum_{k=1}^s \Xi_k^2$ be an {\it
  elliptic} operator.

The good news is that we can get an answer for the expected
correlation functions $\left<\Omega\right>$ and $\left<R_2\right>$ by
taking $\epsilon(\hbar )={\rm const} \times \hbar^{\alpha}$ where
$\alpha$ is a number between 0 and 1 (and we may choose $\alpha =1/2$,
say). This dependence of the variances on $\hbar$ means that, roughly
speaking, all members of the ensemble have the {\it same} classical
limit. In other words, the ensemble we use is ``quantum'', and we
could say that what we are averaging over are different quantizations
of the same classical map.

Now, let the classical map be mixing, and arrange for the quantized
map to have no unitary or anti--unitary symmetries. We will argue that
the expected correlation functions in this case have the following
semiclassical limits:
       \begin{eqnarray}
         \lefteqn{\lim_{N\to\infty}\left<\Omega\left( e^{ic_{+0}/N},
         e^{ic_{+1}/N};e^{-ic_{-0}/N},e^{-ic_{-1}/N}\right)\right> =}
         \nonumber \\
         & & 1-\frac{(c_{+0}-c_{+1})(c_{-0}-c_{-1})}{(c_{+0}-c_{-0})
         (c_{+1}-c_{-1})}\left( 1-e^{i(c_{+1}-c_{-1})}\right) \;,
         \label{rmt}
       \end{eqnarray}
and
          \[
          \lim_{N\to\infty}\left< R_2(2\pi x/N)\right> =
          \delta (x)-\sin^2{(\pi x)}/(\pi x)^2\;.
          \]
To put this result into context, we observe that our expressions for
$\left<\Omega\right>$ and $\left< R_2\right>$ coincide with those for
a random unitary matrix in the large--$N$ limit.  What we propose,
then, is a {\it precise} formulation of the BGS conjecture: random
matrix theory applies to chaotic maps {\it on average} over a quantum
ensemble of the kind specified. More specifically, although there
exist counterexamples such as the cat maps, these are a set of measure
zero in parameter space, and the random matrix result does obtain on
averaging.

\section{Supersymmetric formula for maps (I)}
\label{sec:formula_1}

In what follows we shall construct a novel and useful integral
representation of the correlator $\Omega$. Our first step will be to
express the determinants ${\rm Det}(1-\gamma U_N)^{\pm 1}$ as traces
over Fock space.  To that end, let ${\cal F}_N$ be the
$2^N$--dimensional fermionic Fock space that is generated by acting
with creation operators $f^{\dagger}_n\; (n=1,2,\ldots ,N)$ on a
particle vacuum $\left| 0\right>$. The operators $f^{\dagger}_n$ and
their adjoints $f_n^{\phantom{\dagger}}$ satisfy the canonical
anticommutation relation for fermions: $f^{\dagger}_n
f_{n'}^{\phantom{\dagger}}+ f_{n'}^{\phantom{\dagger}}
f^{\dagger}_n=\delta_{nn'}^{\phantom{\dagger}}$.

Now we claim that determinants such as those appearing in the numerator 
of the defining expression for $\Omega$ can be written as Fock space
traces:
           \[
           {\rm Det}(1-U_N^{\phantom{\dagger}})=
           {\rm Tr}_{{\cal F}_N}^{\phantom{\dagger}}(-1)^{N_F}_{\phantom{n}}
           \exp{\sum\limits_{n,n'=1}^Nf^{\dagger}_n
           (\ln{U_N})_{nn'}^{\phantom{\dagger}}f_{n'}^{\phantom{\dagger}}}
           \]
where $N_F^{\phantom{\dagger}}=
\sum_n^{\phantom{\dagger}}f^{\dagger}_nf_n^{\phantom{\dagger}}$ 
is the fermion number. To prove that 
claim, we may assume the matrix $U_N$ to be diagonal as both sides of the
equation are invariant under a change of the single particle basis.
Put $(U_N)_{nn'}=\lambda_n\delta_{nn'}$. Then the above equation is 
verified by the following simple computation:
           \begin{eqnarray*}
           \prod\limits_{n=1}^N(1-\lambda_n) & = &
           \prod\limits_{n=1}^N\left( e^{0\times\ln{\lambda_n}}-
           e^{1\times\ln{\lambda_n}}\right)\cr
           & = & \prod\limits_{n=1}^N{\rm Tr}_{{\cal F}_1}^{\phantom{\dagger}}
           (-1)^{f^{\dagger}_1f^{\phantom{\dagger}}_1}
           \exp{\left( f^{\dagger}_1f^{\phantom{\dagger}}_1
           \ln{\lambda_n^{\phantom{\dagger}}}\right)}\cr
           & = & {\rm Tr}_{{\cal F}_N}^{\phantom{\dagger}}
           {(-1)}^{N_F}_{\phantom{n}}\exp{\sum\limits_{n=1}^N
           f^{\dagger}_nf_n^{\phantom{\dagger}}
           \ln{\lambda_n^{\phantom{\dagger}}}}\;.
           \end{eqnarray*}
The multivaluedness of the logarithm does not cause any ambiguity here
as $\exp{2\pi if^{\dagger}_nf^{\phantom{\dagger}}_n}=1$.

A similar expression for the inverse of a determinant can be obtained
by substituting bosons for fermions. Let ${\cal B}_N$ be the bosonic
Fock space generated by canonical boson operators $b^{\dagger}_n,\;
b^{\phantom{\dagger}}_n$ with commutation relations $b^{\phantom
  {\dagger}} _{n'} b^{\dagger}_n- b^{\dagger}_nb^ {\phantom{ \dagger}
  }_{n'} = \delta_{nn'}$. For $|\gamma |<1$ we have the identity
          \[
          {\rm Det}(1-\gamma U_N^{\phantom{\dagger}})^{-1}_{\phantom{n}}=
          {\rm Tr}^{\phantom{\dagger}}_{{\cal B}_N}\exp{\sum\limits_{n,n'=1}^N
          b^{\dagger}_n(\ln{\gamma 
          U_N^{\vphantom{\dagger}}}^{\phantom{\dagger}})_{nn'}b_{n'}}\;.
          \]
Again, this is verified by diagonalizing $U_N$ and computing a
single boson trace,
          \[
          {\rm Tr}_{{\cal B}_1}e^{b^{\dagger}_1 b^{\phantom{\dagger}}_1
            \ln\lambda} = e^{0\times\ln{\lambda}}+
          e^{1\times\ln{\lambda}}+\cdots + e^{n\ln{\lambda}}+\cdots =
          (1-\lambda)^{-1}\;.
          \]
The inequality $|\gamma |<1$ is needed in order for the sum on the 
right hand side to converge.

Let us now assemble the various pieces to build a formula for $\Omega$.
There are two determinants in both the numerator and the denominator,
so we introduce two fermionic and two bosonic Fock spaces
${\cal F}^{\pm}_N$ and ${\cal B}^{\pm}_N$, and write
          \begin{eqnarray}
          &&\frac{{\rm Det}(1-\gamma_{+1}^{\vphantom{-1}}U_N^{\vphantom{-1}})
          {\rm Det}(1-\gamma_{-1}^{\vphantom{-1}}U_N^{-1})}
          {{\rm Det}(1-\gamma_{+0}^{\vphantom{-1}}U_N^{\vphantom{-1}})
          {\rm Det}(1-\gamma_{-0}^{\vphantom{-1}}U^{-1}_N)} 
          \nonumber \\
          &=& 
          {\rm Tr}_{{\cal B}^+_N\otimes {\cal F}^+_N\otimes {\cal B}^-_N
          \otimes {\cal F}^-_N} (-1)^{N_F} \exp{\sum\limits_{nn'}} 
          \Big( b^{\dagger}_{+n}(\ln{\gamma_{+0}^{\vphantom{\dagger}}
          U_N^{\vphantom{\dagger}}})_{nn'}^{\vphantom{\dagger}}
          b^{\vphantom{\dagger}}_{+n'}
          \nonumber \\
          &&\hspace{5.5cm} + b^{\dagger}_{-n}(\ln{\gamma_{-0}^{\vphantom{-1}}
          U^{-1}_N})_{nn'}^{\vphantom{\dagger}} b^{\vphantom{\dagger}}_{-n'} 
          \nonumber \\
          &&\hspace{5.5cm} + f^{\dagger}_{+n}(\ln{\gamma_{+1}^{\vphantom{-1}}
          U_N})^{\vphantom{\dagger}}_{nn'} f^{\vphantom{\dagger}}_{+n'}
          \nonumber \\
          &&\hspace{5.5cm} + f^{\dagger}_{-n}(\ln{\gamma_{-1}^{\vphantom{-1}}
          U^{-1}_N})^{\vphantom{\dagger}}_{nn'}f^{\vphantom{\dagger}}_{-n'} 
          \Big) \;.
          \end{eqnarray}
Here $N_F^{\vphantom{\dagger}} = \sum_{n=1}^N\left( f^{\dagger}_{+n} 
  f^{\vphantom{\dagger}}_{+n} + f^{\dagger}_{-n} f^{\vphantom{
      \dagger}}_{-n} \right)$, and convergence of the sums requires
$|\gamma_{\pm 0}|<1$.  To obtain $\Omega$ we need to replace $U_N$ by
$e^{i\theta}U_N$ and integrate over $\theta$. If $N_{\pm}^{\vphantom{
    \dagger}} = \sum_n\left( b^{\dagger}_{\pm n} b^{\vphantom{
      \dagger}}_{\pm n} + f^{\dagger}_{\pm n} f^{\vphantom{
      \dagger}}_{\pm n} \right)$ counts the number of retarded (+) or
advanced (--) particles, this produces an extra factor
  \[
  \int\limits_0^{2\pi}\frac{d\theta}{2\pi}(e^{i\theta})^{N_+ - N_-}
  =: P
  \]
under the trace. The operator $P$ gives unity when applied to a state
$\left|\psi\right>$ with $(N_+-N_-)\left|\psi\right> =0$, and zero
otherwise, and thus projects the super Fock space ${\cal B}^+_N\otimes
{\cal F}^+_N\otimes {\cal B}^-_N\otimes {\cal F}^-_N$ onto the
subspace with equal retarded and advanced particle numbers.  Hence, on
writing $\sum_{nn'} b^{\dagger}_{+n} (\ln{\gamma_{+0}^{\vphantom
    {\dagger}} U^{\vphantom{\dagger}}_N})^{\vphantom{\dagger}}_{nn'}
b^{\vphantom{\dagger}}_{+n'} =:b^{\dagger}_+ \ln{ (\gamma^{\vphantom
    {\dagger}}_{+0} U^{\vphantom{\dagger}}_N)} b^{\vphantom{\dagger}}_+$
etc. to simplify the notation, we obtain
           \begin{eqnarray}
           \lefteqn{
           \Omega (\gamma_{+0},\gamma_{+1};\gamma_{-0},\gamma_{-1})=}
           \hspace{7.3cm}
           \nonumber \\
           {\rm Tr}^{\vphantom{\dagger}}_{N_+=N_-}
           (-1)^{N_F}_{\vphantom{+}}\exp \Big( b^{\dagger}_+
           \ln{(\gamma^{\vphantom{\dagger}}_{+0}U^{\vphantom{\dagger}}_N)}
           b^{\vphantom{\dagger}}_+ & + &b^{\dagger}_- \ln{(
           \gamma^{\vphantom{-1}}_{-0}U^{-1}_N)} b^{\vphantom{\dagger}}_-
           \nonumber \\
           + f^{\dagger}_+ \ln{(\gamma^{\vphantom{-1}}_{+1}
           U_N)} f^{\vphantom{\dagger}}_+ & + &
           f^{\dagger}_- \ln{(\gamma^{\vphantom{-1}}_{-1}U^{-1}_N)}
           f^{\vphantom{\dagger}}_- \Big) \;.
           \label{trace}
           \end{eqnarray}

The final step in the construction of the supersymmetric formula for
$\Omega$ is to pass from the trace over super Fock space to a Berezin
integral, by the introduction of generalized coherent states. This
step is perhaps unfamiliar and deserves a certain amount of
explanation.

Let me first illustrate the basic idea at the example of $\gamma_{+0}
= \gamma_{-0} = 0$, in which case the bosonic degrees of freedom are
absent. Thus, we start from the relation
          \[
          \Omega (0,\gamma^{\vphantom{\dagger}}_{+1};0,
          \gamma^{\vphantom{\dagger}}_{-1}) =
          {\rm Tr}^{\vphantom{\dagger}}_{{\cal F}^+_N\otimes 
          {\cal F}^-_N}P^{\vphantom{\dagger}}_F 
          \exp{\left(
          f^{\dagger}_+ \ln{(\gamma^{\vphantom{\dagger}}_{+1}
          U^{\vphantom{\dagger}}_N)} f^{\vphantom{\dagger}}_+ +
          f^{\dagger}_- \ln{(\gamma^{\vphantom{\dagger}}_{-1}
          U^{-1}_N)} f^{\vphantom{\dagger}}_-\right)}
          \]
where $P_F = (2\pi )^{-1}_{\vphantom{n}}\int_0^{2\pi} {\rm d} \theta
\exp{i\theta \sum_n( f^{\dagger}_{+n}f^{\vphantom{\dagger}}_{+n}-
  f^{\dagger}_{-n}f^{\vphantom{\dagger}}_{-n})}$. Consider then the BCS
(or pair) coherent states
          \[
          \left| Z\right> = 
          \exp{\left(\sum_{n,n'=1}^N f^{\dagger}_{+n}
          Z^{\vphantom{\dagger}}_{nn'}f^{\dagger}_{-n'}\right)}
          \left| 0\right>
          \]
with complex amplitudes $Z_{nn'}$. Being made from pairs of one
retarded and one advanced particle, these states satisfy the
constraint $P_F\left| Z\right> =\left| Z\right>$. Moreover, they enjoy
the key property of providing a resolution of the projector $P_F$:
          \[
          P_F=\int\limits_{{\Bbb C}^{N\times N}}d\mu_N(Z,\bar Z)
          \left| Z\right> \left< Z\right|\,,
          \]
where $d\mu_N (Z,\bar Z)={\rm const}\times {\rm Det} (1+Z^\dagger
Z)^{-2N-1} \prod_{n,n'=1}^N {\rm d}Z_{nn'}\, {\rm d}\bar Z_{nn'}$.
The proof proceeds via Schur's lemma, by demonstrating that $\int
d\mu_N(Z,\bar Z) \left| Z\right> \left< Z\right|$ commutes with all of
the operators $f^{\dagger}_{+n}f^{\vphantom{\dagger}}_{-n'}$ and
$f^{\dagger}_{-n}f^{\vphantom{\dagger}}_{+n'}\; (n,n'=1,\ldots ,N)$.  A
standard computation on coherent states \cite{perelomov} yields
          \begin{eqnarray*}
          \left< Z\right|\exp{\sum\limits_{nn'}\left(
          f^{\dagger}_+
          \ln{(\gamma^{\vphantom{\dagger}}_{+1}U^{\vphantom{\dagger}}_N)}
          f^{\vphantom{\dagger}}_+ +f^{\dagger}_- 
          \ln{(\gamma^{\vphantom{\dagger}}_{-1}U^{-1}_N)} 
          f^{\vphantom{\dagger}}_-\right)}\left| Z\right>\\
          = {\rm Det}(1+\gamma^{\vphantom{\dagger}}_{+1}
          \gamma^{\vphantom{\dagger}}_{-1}Z^{\dagger}
          U^{\vphantom{\dagger}}_NZU^{-1}_N)\;.
          \end{eqnarray*}
Hence, by inserting the coherent state resolution of $P_F$ into 
the expression for $\Omega (0,\gamma_{+1};0,\gamma_{-1})$, we arrive at
the formula
          \[
          \Omega (0,\gamma_{+1}^{\vphantom{\dagger}};0,
          \gamma^{\vphantom{\dagger}}_{-1})=\int\limits_{{\Bbb C}^{N\times N}}
          d\mu_N^{\vphantom{\dagger}}(Z,\bar Z)\,
          {\rm Det}\left( 1+\gamma^{\vphantom{\dagger}}_{+1}
          \gamma^{\vphantom{\dagger}}_{-1}Z^{\dagger}
          U^{\vphantom{\dagger}}_NZU^{-1}_N\right)\;.
          \]
Our aim now is to extend this formula to allow for a nonvanishing
first and third argument of $\Omega$, in which case the Fock space
expression involves both fermions and bosons. This requires a
supergeneralization of the notion of coherent states and their
integration measure. The purpose of the next section is to review some
of the basic mathematics needed.  We will return to the task of
expressing $\Omega$ as a coherent state integral in
Sect.~\ref{sec:formula_2}.

\section{Basic notions of analysis on supermanifolds}
\label{sec:basics}

Familiarity with the elementary theory of differentiable manifolds is
assumed.  In the present section the summation convention is in force.

Let $\Lambda {\Bbb R}^q$ be the Grassmann algebra over ${\Bbb R}$ with
$q$ generators $\psi^1,\psi^2,\ldots ,\psi^q$. By definition, the
Grassmann generators anticommute: $\psi^i\psi^j+\psi^j\psi^i=0$.  They
are known to physicists from their use in the construction of path
integrals for fermions. Note in particular $\psi^i\psi^i =
-\psi^i\psi^i = 0$.

Now pick some domain $U\subset{\Bbb R}^p$. A map $F$ from $U$ into 
the Grassmann algebra,
          \begin{eqnarray*}
          F:\quad U & \to & \Lambda {\Bbb R}^q \\
          a & \mapsto & F_0(a)+F_i(a)\psi^i
          +F_{i_1i_2}(a)\psi^{i_1}\psi^{i_2}
          + \cdots + F_{i_1\cdots i_q}(a)\psi^{i_1}\cdots \psi^{i_q}\,,
          \end{eqnarray*}
will be referred to as a {\it superfunction} (or simply a function)
on $U$. $F_0$ is called the {\it number part} of $F$. We distinguish
between {\it even} and {\it odd} superfunctions (the former are even
and the latter are odd in the Grassmann generators). This distinction
endows the algebra of superfunctions with a ${\Bbb Z}_2$--{\it
  grading}.

We can now describe what is meant by a $(p,q)$--dimensional
real--analytic supermanifold in the sense of Berezin, Kostant and
Leites \cite{bkl}.  Take a (real--analytic) manifold $M$ of dimension
$p$ and cover it by domains $U_1,U_2,\ldots $. Let $p$ even and $q$
odd coordinate superfunctions be given on each domain. [For the
purpose of illustration, we denote these by
          \begin{eqnarray*}
          & &  x^1,\ldots ,x^p;\; \xi^1,\ldots ,\xi^q\quad {\rm on}\; 
          U_1\,, \\
          {\rm and\; by} & & y^1,\ldots ,y^p;\; \eta^1,\ldots ,\eta^q 
          \quad {\rm on}\; U_2\;.]
          \end{eqnarray*}
Assume that transition functions exist on overlapping domains and are
consistent with the ${\Bbb Z}_2$--grading. [For example, on
$U_1\bigcap U_2$ this means that there are relations
          \begin{eqnarray*}
          & & y^i=f^i(x,\xi )=f^i_0(x)+f^i_{kl}(x)\xi^k\xi^l+\cdots\,,\\
          & & \eta^j=\varphi^j(x,\xi )=\varphi^j_k(x)\xi^k+
          \varphi^j_{klm}(x) \xi^k\xi^l\xi^m +\cdots
          \end{eqnarray*}
(and the corresponding inverse relations) with $f^i_0,\; f^i_{kl},\;
\varphi^j_k,\;\varphi^j_{klm}$ etc. being functions on a subset of
${\Bbb R}^p$.] If the transition functions are analytic functions, we
call the algebra of superfunctions on $M$ generated by the coordinate
superfunctions a {\it real--analytic supermanifold} of {\it dimension}
$(p,q)$. Complex--analytic supermanifolds are defined in an analogous
manner (just replace $\Lambda {\Bbb R}^q$ by $\Lambda {\Bbb C}^q$, the
Grassmann algebra over ${\Bbb C}$ with $q$ generators, and require the
transition functions to be holomorphic).  The manifold $M$ is called
the {\it base} of the supermanifold.

Of course, a proper definition avoids any reference to specific
coordinates, and in a more mathematical exposition \cite{berezin} one
would define a supermanifold to be a ``sheaf of algebras of
superfunctions ...''.  However, the coordinate description given here
is good enough for our purposes.

The standard example of a complex--analytic supermanifold is ${\Bbb
  C}{\rm P}^{1|1}$.  Let $z^1,z^2$ be canonical coordinates of ${\Bbb
  C}^2$, and consider the algebra of superfunctions: ${\Bbb C}^2
\setminus \{ 0\}\to \Lambda{\Bbb C}^1$, with the complex Grassmann
generator of $\Lambda{\Bbb C}^1$ denoted by $\zeta^1$.  Then focus on
the subalgebra of superfunctions which are homogeneous of degree zero
in these generators, i.e. are invariant under rescalings $z^1\to cz^1,
\; z^2\to cz^2,\; \zeta^1\to c\zeta^1$ with $c\in{\Bbb C}\setminus\{
0\} $. This algebra is what is called the complex projective
superspace ${\Bbb C}{\rm P}^{1|1}$. Its base is the ordinary complex
projective space ${\Bbb C}{\rm P}^1$. Because the null element has been
removed from ${\Bbb C}^2$, at least one of the two coordinates $z^1,z^2$ 
is nonzero. Let $U_1$ be the domain where $z^1\not= 0$.  Then homogeneity
means
           \[
           f(z^1,z^2,\zeta^1)=f(1,z^2/z^1,\zeta^1/z^1)\,,
           \]
so that $f$ is a function of two variables $z:=z^2/z^1$ and $\zeta
:=\zeta^1/z^1$, which are taken to be the coordinate (super)functions
on $U_1$. Similarly on $U_2$, defined by $z^2\not= 0$, we have
           \[
           f(z^1,z^2,\zeta^1)=f(z^1/z^2,1,\zeta^1/z^2)\,,
           \]
and here we put $z' := z^1/z^2$ and $\zeta':=\zeta^1/z^2$.
The transition functions on $U_1\bigcap U_2$ are then given by
           \[
           z'=\frac{1}{z},\quad \zeta '=\frac{\zeta}{z}\;.
           \]
The domains $U_1$ and $U_2$ cover the base ${\Bbb C}{\rm P}^1$, and
the transition functions are holomorphic, so ${\Bbb C}{\rm P}^{1|1}$
is a complex--analytic supermanifold, the complex dimension being
(1,1).

Of the many structures that exist on supermanifolds, the most
important one for our purposes is the Berezin integral.  To define it,
we must first introduce the notion of {\it Berezin form}, also called
superintegration measure or integral superform. If ${\cal A}$ denotes
the algebra of superfunctions of our supermanifold and $\Omega^* M$
the space of volume forms on the base $M$, a Berezin form $\omega$ is
a linear and local map
           \begin{eqnarray*}
           \omega:\quad {\cal A} & \to & \Omega^{\ast}M\,,\\
           f & \mapsto & \omega [f]\,,
           \end{eqnarray*}
i.e. a linear and local rule for converting superfunctions into volume
forms on $M$.  This process of conversion is sometimes called the
``Fermi integral'' or ``integration over the Grassmann variables''.

For example, an interesting class of Berezin forms on ${\Bbb C} {\rm
  P}^{1|1}$ is
           \[
           \omega^{(n)} = {\rm d}z\wedge {\rm d}\bar z \circ 
           \frac{\partial^2}{\partial\zeta\partial \bar\zeta} \circ
           (1+\bar zz+\bar\zeta\zeta )^{-n}
           \]
for $n\in N$. The meaning hereof is this: to compute $\omega^{(n)}
[f]$, one first multiplies the superfunction $f$ by $(1 + \bar zz+
\bar\zeta\zeta)^{-n}$, then one takes two derivatives $\partial^2 /
\partial\zeta\partial\bar\zeta$ (thereby removing the Grassmann
generators), and finally one multiplies by ${\rm d}z\wedge {\rm d}\bar
z$ to obtain a two--form on ${\Bbb C}{\rm P}^1$.  To illustrate, we
take
           \[
           f=f_0(z,\bar z)+f_{01}(z,\bar z)\bar\zeta +
           f_{10}(z,\bar z)\zeta +f_{11}(z,\bar z)\bar\zeta\zeta\,,
           \]
in which case we get
           \[
           \omega^{(n)} [f] = 
           \left(\frac{f_{11}(z,\bar z)}{(1+\bar zz)^n} -
           n\frac{f_{00}(z,\bar z)}{(1+\bar zz)^{n+1}}\right)
           {\rm d}z\wedge {\rm d}\bar z \;.
           \]
A special role is played by the Berezin form $\omega^{(1)}$.
Converting $\omega^{(1)}$ from one chart to another by using the
transition functions $z = 1/z'$ and $\zeta = \zeta'/z'$, one finds
that $\omega^{(1)}$ operates by the same expression in both charts.
(Moreover, $\omega^{(1)}$ is invariant under an action of the Lie
supergroup ${\rm SU}(2|1)$ on ${\Bbb C}{\rm P}^{1|1}$.)

The definition of Berezin forms immediately leads to a notion of
integration on supermanifolds called the Berezin integral, or
superintegral. Since a Berezin form $\omega$ converts a superfunction
$f$ into a volume form $\omega [f]$ on $M$, the natural thing to do is
to integrate $\alpha :=\omega [f]$ in the ordinary fashion to produce
the number $\int_M \alpha$. Thus a Berezin integral is defined
to be a composite map
           \[
           f\mapsto\omega [f]\mapsto\int_M \omega [f]\,,
           \]
taking superfunctions into the complex numbers. Note that this 
definition decrees the Berezin integral to be a two--step process:
first the ``integration'' (or, rather, differentiation) of the
Grassmann generators is done, and afterwards the differential form
$\omega [f]$ is integrated in the usual sense.  The advanced user
occasionally finds it convenient to deviate from this rule and perform
part of the ordinary integrals first. However, in case of doubt the
definition recited \cite{berezin} above is the one to go back to.

Ordinary integration is coordinate independent, i.e. one has the
option of changing integration variables by the substitution rule.
It turns out that this option also exists for superintegrals,
albeit with a certain caveat. To describe the relevant statement,
which is due to Berezin, we introduce the short hand notation
  \[
  D(x,\xi):= {\rm d}x^1\wedge\ldots \wedge {\rm d}x^p
  \frac{\partial}{\partial\xi^1}\cdots
  \frac{\partial}{\partial\xi^q}\,,
  \]
and consider a change of variables (consistent with the ${\Bbb
  Z}_2$-grading) from $x,\xi$ to $y,\eta$ by
          \begin{eqnarray*}
          y^i & = & f^i(x^1,\ldots ,x^p;\xi^1,\ldots ,\xi^q)
          \qquad (i=1,...,p)\,,\\
          \eta^j & = & \varphi^j(x^1,\ldots ,x^p;\xi^1,\ldots ,\xi^q)
          \qquad (j=1,...,q)\;.
          \end{eqnarray*}
The role of the Jacobian of ordinary analysis is taken by the
{\it Berezinian},
          \[
          {\rm Ber}\left(\frac{y,\eta}{x,\xi}\right) :={\rm SDet}
          \left( \begin{array}{ll}
          \partial f^i/\partial x^j & \partial f^i/\partial\xi^j\\ 
          \partial\varphi^i/\partial x^j & \partial\varphi^i/\partial\xi^j 
          \end{array}\right)\,,
          \]
where the {\it superdeterminant} SDet of a supermatrix 
$\left(\begin{array}{cc} A & B\\ C & D\end{array}\right)$ (with the 
matrix elements of $A,D$ being even and those of $B,C$ odd) is defined by
          \[
          {\rm SDet}\left(\begin{array}{cc} A & B\\ C & D\end{array}\right)
          =\frac{{\rm Det}(A-BD^{-1}C)}{{\rm Det}D}=
          \frac{{\rm Det}A}{{\rm Det}(D-CA^{-1}B)}\;.
          \]
Berezin's theorem \cite{berezin} states that the substitution rule is 
valid, i.e.
          \[
          \int {D}(y,\eta )f=\int{D}(x,\xi ) \; {\rm Ber}\left(
          \frac{y,\eta}{x,\xi}\right)f\,,
          \]
provided that $f$ is compactly supported.

The condition of compact support, unfamiliar from ordinary analysis,
invites some explanation. For an instructive example \cite{rothstein},
consider the noncompact supermanifold of functions: $]0,1[ \to \Lambda
{\Bbb R}^2$, with canonical coordinates $x,\xi^1,\xi^2$.  If we choose
to integrate with the Berezin form ${\rm d}x \; \partial^2 / \partial
\xi^2 \partial\xi^1$, a superfunction $f$ that depends only on the
combination $x+\xi^1\xi^2$ has the Berezin integral
          \[
          \int\limits_0^1 {\rm d}x 
          \frac{\partial^2}{\partial\xi^2\partial\xi^1}
          f(x+\xi^1\xi^2) = \int\limits_0^1 {\rm d}x f'(x) = f(1)-f(0)\;.
          \]
On the other hand, if we made a change of integration variables
          \[
          y=x+\xi^1\xi^2,\quad \eta^1=\xi^1,\quad \eta^2=\xi^2\,,
          \]
which is easily seen to have unit Berezinian, we would get
          \[
          \int\limits_0^1 {\rm d}x 
          \frac{\partial^2}{\partial\xi^2\partial\xi^1}
          f(x+\xi^1\xi^2)
          \; {\buildrel ?? \over =} \;
          \int\limits_0^1 {\rm d}y
          \frac{\partial^2}{\partial\eta^2\partial\eta^1}f(y)=0\;.
          \]
Because the Grassmann derivatives $\partial^2/\partial \eta^2 \partial
\eta^1$ have nothing to act on, the right hand side vanishes although
for equality with the left hand side, it ought to be $f(1)-f(0)$.
However, both sides vanish, and the discrepancy disappears, if we take
$f$ to be compactly supported (which implies $f(0)=f(1)=0$), as is
required in order for Berezin's theorem to apply. An alternative way
of fixing the problem is to pass from the noncompact interval $]0,1[$
to the compact circle ${\rm S}^1$, by imposing periodic boundary
conditions on $x$, so that $f(1)-f(0)=0$.

The above example signals a general complication, which one has to
confront when changing variables in a superintegral. While ordinary
volume forms transform simply by the Jacobian, the transformation law
for Berezin forms is {\it not} given only by the Berezinian, but
involves an additional, ``anomalous'' term $\beta$:
          \[
          D(y,\eta) = D(x,\xi) \;
          {\rm Ber}\left(\frac{y,\eta}{x,\xi}
          \right) +\beta\,,
          \]
whenever some of the even coordinates are shifted by nilpotents. 
In our simple example,
          \[
          {D}(y,\eta )f = {D}(x,\xi )f+
          {\rm d}x \frac{\partial}{\partial x}f
          \Big|_{\xi^1=\xi^2=0}\;.
          \]
The anomaly $\beta$ always has the property of being exact (in the
sense of differential forms), i.e. $p[f]={\rm d}\alpha [f]$ where
${\rm d}\alpha [f]$ is the exterior derivative of some $(p-1)$--form
$\alpha [f]$. Therefore, by Stokes' theorem the anomaly can be
integrated and converted into an integral over the boundary of the
base of the supermanifold (or the boundary of the chart used, if there
exist coordinate singularities). Rothstein \cite{rothstein} has given
an explicit expression for the anomaly in terms of the vector field
generating the transformation from $x,\xi$ to $y,\eta$.

\section{Supersymmetric formula for maps (II)}
\label{sec:formula_2}

After this brief tour of some basic supermathematics, we return to the
task of setting up a supersymmetric formula for the pair correlations
of quantized symplectic maps. We left off at Eq.~(\ref{trace})
expressing $\Omega$ as a trace over $S$, the subspace of super
Fock space ${\cal B}^+_N\otimes {\cal F}^+_N\otimes {\cal B}^-_N\otimes
{\cal F}^-_N$ determined by the condition $N_+=N_-$ (equal number of
retarded and advanced particles).

As was motivated at the example of the purely fermionic case in
Sect.~\ref{sec:formula_1}, our goal is now to trade the trace over $S$
for an integral over generalized coherent states.  To that end, we
observe that every Fock state $\left|\psi\right>$ satisfying
$(N_+-N_-)\left|\psi\right> = 0$ can be obtained by repeatedly acting
on the vacuum with the pair creation operators
$b^\dagger_{+n}b^\dagger_{-n'}$, $b^\dagger_{+n}f^\dagger_{-n'}$,
$f^\dagger_{+n}b^\dagger_{-n'}$, and $f^\dagger_{+n}f^\dagger_{-n'}$
($n,n'=1,\ldots ,N$).  This leads us to consider coherent states of
the pairing form
            \[
            \big| Z\big> :=\exp{\sum\limits_{nn'}\left(
            b^{\dagger}_{+n}Z^{\rm BB}_{nn'}b^{\dagger}_{-n'}+
            b^{\dagger}_{+n}Z^{\rm BF}_{nn'}f^{\dagger}_{-n'}+
            f^{\dagger}_{+n}Z^{\rm FB}_{nn'}b^{\dagger}_{-n'}+
            f^{\dagger}_{+n}Z^{\rm FF}_{nn'}f^{\dagger}_{-n'}
            \right)}\big| 0\big>\,,
            \]
where mathematical consistency requires taking complex numbers for the 
matrix elements $Z^{\rm BB}_{nn'},\; Z^{\rm FF}_{nn'}$ and Grassmann
generators for $Z^{\rm BF}_{nn'},\; Z^{\rm FB}_{nn'}$.
It is convenient to assemble these into a supermatrix
            \[
            Z=\left( \begin{array}{cc} Z^{\rm BB} & Z^{\rm BF}\\
            Z^{\rm FB} & Z^{\rm FF} \end{array}\right)\;.
            \]
Then a short hand notation for the pair coherent states is
            \[
            \big| Z\big> =\exp{\left( c^{\dagger}_+
             Z c^{\dagger}_-\right) }\big| 0\big>
            \]
where the creation operators $c^{\dagger}_{\pm}$ stand for both
$b^{\dagger}_{\pm}$ and $f^{\dagger}_{\pm}$. We will also need
the dual of a pair coherent state
            \[
            \big<\tilde Z\big| :=\big< 0\big|\exp{\left(
            c_- \sigma\tilde Z c_+\right) },\quad
            \tilde Z=\left(
            \begin{array}{cc} \tilde Z^{\rm BB} & \tilde Z^{\rm BF}\\
            \tilde Z^{\rm FB} & \tilde Z^{\rm FF}\end{array}\right),
            \quad \sigma =\left(\begin{array}{cc} {1}_N & 0\\
            0 & -{1}_N\end{array}\right)\;.
            \]
In \cite{mrz_circular} it was shown that, with a suitable choice of
Berezin form $D(Z,\tilde Z)$, the pair coherent states provide a
resolution of the projector $P$,
            \[
            P=\int\limits_0^{2\pi}\frac{d\theta}{2\pi}\exp{i\theta
            (N_+-N_-)} = \int D(Z,\tilde Z) \;            
            \big<\tilde Z\big| Z\big>^{-1} \;
            \big| Z\big> \big< \tilde Z\big| \,,
            \]
where the domain of integration is fixed by the conditions
            \[
            \tilde Z^{\rm FF}=-{Z^{\rm FF}}^{\dagger} \;, \quad 
            \tilde Z^{\rm BB}= +{Z^{\rm BB}}^{\dagger} \;, \quad
            {\rm and} \quad 1-{Z^{\rm BB}}^{\dagger}Z^{\rm BB}>0 \;.
            \]
This resolution follows from Schur's lemma, once it has been

demonstrated that the right hand side commutes with all bilinears in
Fock operators that commute with the projector $P$. The crucial
relation needed for the latter statement is the invariance of the
Berezin integral under a left translation through $g = \left(
  \begin{array}{cc} A & B\\ C & D \end{array}\right) \in {\rm GL}(2N|2N)$, 
           \begin{eqnarray*}
             \lefteqn{\int D(Z,\tilde Z)f(Z,\tilde Z) =}\\ & & \int
             D(Z,\tilde Z)f\Big( (AZ+B)(CZ+D)^{-1}, (C+D\tilde
             Z)(A+B\tilde Z)^{-1}\Big) \;,
           \end{eqnarray*}
which determines the Berezin form $D(Z,\tilde Z)$ to be the invariant
superintegration measure on Efetov's model space II (or, using the
terminology of \cite{mrz_suprev}, on a Riemannian symmetric superspace
of type AIII/AIII).  The invariant Berezin form $D(Z,\tilde Z)$ turns
out to be flat as a result of cancellations due to supersymmetry:
           \[
           D(Z,\tilde Z) = \prod_{n,n'}
           {\rm d}Z_{nn'}^{\rm BB}
           {\rm d}\tilde Z_{nn'}^{\rm BB}
           {\rm d}Z_{nn'}^{\rm FF}
           {\rm d}\tilde Z_{nn'}^{\rm FF}
           {\partial^4 \over 
             \partial Z_{nn'}^{\rm BF}
             \partial\tilde Z_{nn'}^{\rm BF}
             \partial Z_{nn'}^{\rm FB}
             \partial\tilde Z_{nn'}^{\rm FB}} + ...
           \]
modulo boundary anomalies, which are indicated by the dots.  The
normalization of $D(Z,\tilde Z)$ is fixed by $1=\big< 0\big| P\big|
0\big> =\int D(Z,\tilde Z)\big<\tilde Z\big| Z\big>^{-1}$.

Given the coherent state resolution of the projector $P$,
the trace of any operator $O$ on Fock space can be converted
into a Berezin integral:
          \[
          {\rm Tr}_{N_+=N_-}(-1)^{N_F}O ={\rm STr} PO 
          =\int D(Z,\tilde Z) \big<\tilde Z\big| 
            O\big| Z\big> / \big<\tilde Z\big| Z\big> \;.
          \]
Application to Eq.~(\ref{trace}), followed by a computation of $\big<
\tilde Z\big| Z\big>$ and $\big<\tilde Z\big|\ldots\big| Z\big>$,
results in the desired formula for $\Omega$:
          \begin{equation}
          \Omega (\gamma_+^{\vphantom{1}};\gamma_-^{\vphantom{1}})
          =\int D(Z,\tilde Z) \; {\rm SDet}(1-\tilde ZZ) \,
          {\rm SDet}^{-1}\Big( 1-\tilde Z(\gamma_+^{\vphantom{1}}
          \otimes U_N^{\vphantom{1}})
          Z(\gamma_-^{\vphantom{1}}\otimes U^{-1}_N)\Big)\;.
          \label{susy_formula}
          \end{equation}
Here $\gamma_{\pm}={\rm diag}(\gamma_{\pm 0},\gamma_{\pm 1})$ are
diagonal $2\times 2$ matrices, and $\gamma_+\otimes U_N$ is meant to
be a supermatrix:
          \[
          \gamma_+\otimes U_N =\left(\begin{array}{cc}
          \gamma_{+0}U_N & 0\\ 0 & \gamma_{+1}U_N\end{array}\right)\;.
          \]
A closely related derivation of the formula for $\Omega$ is provided
by the so--called color--flavor transformation (\cite{mrz_circular}
and \cite{mrz_icmp97}).  The two--level correlation function $R_2$ 
follows from $\Omega$ by taking two derivatives, as before.

\section{Semiclassical limit?}
\label{sec:semiclassics}

So far, we have achieved no more (and no less) than an exact
reformulation of the original problem. As should be clear from its
derivation, the supersymmetric formula for the correlator $\Omega$
applies to {\it any} unitary matrix $U_N$, independently of whether
this matrix arises from quantization of a map or not.  In order to go
further and extract nonempty information from our formalism, we need
to exploit the fact that $U_N$ has a semiclassical limit and, in
particular, we will have to make a distinction between chaotic and
integrable systems.

To prepare these steps, we write $\Omega = \int D(Z,\tilde Z) e^{-S}$
where, borrowing the terminology from quantum field theory, the
function 
      \begin{equation} 
        S = - {\rm STr} \ln (1 - \tilde Z Z) + {\rm STr} \ln 
        \left( 1 - \tilde Z (\gamma_+ \otimes U_N) Z (\gamma_-
        \otimes U_N^{-1}) \right) 
        \label{action_S} 
       \end{equation} 
will be called the ``action functional''.  Now recall that the ${\rm
  BB}$-- and ${\rm FF}$--blocks of $Z$ are complex $N \times N$
matrices.  Viewing them as linear operators on the Hilbert space
${\cal H}_N$, we might say that they are similar to the density matrix
of a quantum mechanical state.  And, in fact, an expression like
${Z^{\rm BB}}' := U_N Z^{\rm BB} U_N^{-1}$, which occurs under the
second logarithm in $S$, can be interpreted as being the ``density
matrix'' $Z^{\rm BB}$ evolved by one (inverse) time step.  One may
therefore be tempted \cite{a3s} to transform to a Wigner
representation, postulate a semiclassical limit for $Z$ and subject
the action functional to a semiclassical expansion.  However, such an
expansion is {\it entirely uncontrolled} in the present context and
is, in fact, false.  Unfortunately, a lot of confusion has been
created among the community (including myself) concerning this point.
Let me therefore change the style of presentation and give a detailed
exposition of the issues as I see them.  The reader is warned that,
given the current level of understanding, the following has to be
somewhat qualitative.

If $A$ is a linear operator (on some quantum mechanical Hilbert space)
with kernel $A(q,q')$ in, say, the position representation, one defines
the Wigner transform (or Weyl symbol) of $A$ by 
       \[
       \sigma_A(q,p) = 
       \sum_{q'} A( q+q'/2 , q-q'/2) \; e^{ipq'/\hbar}
       \]
whenever the transform exists.  (In our case there is a technical
complication due to the fact that the ``position'' on a compact phase
space is not globally defined, but this is a peripheral issue and we
are not going to worry about it here.)  Now if $A$ and $B$ are
pseudodifferential operators with Wigner transforms $\sigma_A(q,p)$
and $\sigma_B(q,p)$, their operator product has a semiclassical
expansion
       \begin{equation}
       \sigma_{AB} = \sigma_A \sigma_B + (\hbar/2i) \{ \sigma_A , 
       \sigma_B \} + {\cal O}(\hbar^2) ,
       \label{moyal}
       \end{equation}
where $\sigma_A \sigma_B$ is a product of functions, and the curly
brackets denote the Poisson bracket.

The question now is whether this kind of expansion applies to products
of the supermatrix $Z$.  To be sure, there is nothing that prevents us
from transforming $Z$ to a Wigner-like representation, but are we
really allowed to make a semiclassical expansion of the action
functional?!  The question is a valid one, for the semiclassical
expansion (\ref{moyal}) does require $A$ and $B$ to be {\it
  pseudodifferential operators}.  Without the smoothness property
provided by this condition, the Wigner transform of the operator
product $AB$ does {\it not} separate into the product of Wigner
transforms $\sigma_A \sigma_B$ in the limit $\hbar \to 0$.  For the
standard operators appearing in quantum mechanics (position, momentum,
energy etc.)  the smoothness condition is of course always satisfied.
However, there is {\it no} principle that guarantees smoothness of the
Wigner transform $\sigma_Z$ in the present case.  Indeed, $Z$ is not a
fixed observable but is a {\it variable of integration}.  While a
small fraction of the $Z$ that are integrated over do correspond to
smooth $\sigma_Z$, the vast majority do not. (Recall that the integral
is over all complex supermatrices $Z, \tilde Z$ with the integration
measure being that of a Riemanian symmetric superspace.)  This
invalidates the error estimate in (\ref{moyal}).  One may still
cherish the hope that the {\it dominant} contributions to the
$Z$--integral do come from smooth configurations.  Admittedly, this is
what happens for Feynman's path integral in the limit $\hbar \to 0$,
which is dominated by paths that are extrema of the action functional.
However, the present situation is quite different. {\it There exists
  no saddle--point or stationary--phase or other principle here that
  would favor smoothness} in general.

To recognize the severity of the problem, recall that the unitary
operator $U_N$ has eigenvalues $e^{i\theta_n}$ and eigenfunctions
$\psi_n$, and there exist $N$ of them.  As a result, the conjugation
${\rm Ad}(U_N) : \rho \mapsto U_N^{\vphantom{-1}} \rho U_N^{-1}$ has
$N$ eigendensities $\rho = \psi_n \otimes \bar\psi_n$ with eigenvalues
$e^{i(\theta_n - \theta_n)} = 1$.  To compute the pair correlation
function in the microlocal scaling limit, we need to set $\gamma_\pm =
e^{ic_\pm / N}$ and let $N$ go to infinity.  In this limit the action
functional acquires $N$ {\it zero modes} given by $Z = \sum_{n = 1}^N
C^{(n)} \psi_n \bar\psi_n$, where the $C^{(n)}$ are $2\times 2$
supermatrices and are otherwise arbitrary.  Thus there exist $N$
directions through the point $Z = \tilde Z = 0$ along which the
integrand is {\it exactly neutral} -- a circumstance that surely
causes very--large--amplitude fluctuations about this point.
Moreover, the zero mode densities $\psi_n \bar\psi_n$ are never
smooth.  Indeed, in the integrable limit they are sharply localized on
invariant sets, while in the chaotic limit they are highly irregular
functions that vary on the shortest scales permitted by the
uncertainty principle.  In other words, the action functional
(\ref{action_S}) is unstable with respect to a large number $N$ of
zero modes and none of these has a smooth Wigner transform.  In view
of all this, we had better abandon the hope that the dominant
contributions to the $Z$-integral come from configurations that
possess the smoothness of a pseudodifferential operator.  Quite on the
contrary, the most important configurations are those where $\sigma_Z$
varies with the shortest wave length possible!  As $\hbar$ is lowered,
the dominance of the $N$ zero modes keeps introducing an increasing
number of relevant field configurations that vary on finer and finer
scales. In order for the semiclassical asymptotics (\ref{moyal}) to
set in, we would have to intervene and {\it limit} the variations of
$\sigma_Z$ to wave lengths that are greater than some fixed (i.e.
$\hbar$-independent) minimal scale, and then let $\hbar$ go to zero.
Without such an intervention, the Poisson bracket (and higher) terms
in (\ref{moyal}) fail to be $\hbar$-independent but {\it diverge}
(with increasing powers) as $\hbar\to 0$, so that truncation of the
semiclassical expansion of $S$ is totally unjustified.  Consequently,
any conclusion resulting from the use of this expansion in
(\ref{action_S}) is at risk to be false.

Given the invalidity of the semiclassical step, how can we use the
supersymmetric formula (\ref{susy_formula}) to develop the theory
further?  Not surprisingly, the answer depends on what we are trying
to achieve.  Here we are pursuing no more than the modest goal (modest
from a physicist's perspective) of formulating a precise version of
the BGS conjecture and substantiating it.  As will be shown in the
sequel, this restricted goal offers us the option of postponing the
semiclassical step until the very end of the calculation.

\section{Regularization}
\label{sec:regularization}

Notwithstanding the fact that the formula (\ref{susy_formula}) is
well-defined and, in fact, mathematically rigorous, we are facing a
key difficulty: the existence of $N$ one--parameter groups along which
the integrand lacks stability prevents us from investing semiclassical
input and makes it difficult if not impossible to compute something
from (\ref{susy_formula}) as it stands.  To make any progress at all,
we must first solve this stability problem.  Consider therefore, in
the microlocal scaling limit $\gamma_\pm \to 1$, the quadratic part of
the action functional at $Z = \tilde Z = 0$:
      \[
      S^{(2)} = {\rm STr}\; \tilde Z Z - {\rm STr}\; \tilde Z U_N Z U_N^{-1}
      = {\rm STr}\; \tilde Z \big( 1 - {\rm Ad}(U_N) \big) Z \;,
      \]
which identifies the Hessian at this point as the operator $1 - {\rm
  Ad}(U_N)$.  Now recall that the classical limit of ${\rm Ad}(U_N)$ is
$\chi^*$, the Frobenius--Perron operator of the inverse map
$\chi^{-1}$.  To prepare the discussion of the quantum mechanical
operator ${\rm Ad}(U_N)$ in the limit $N \to \infty$, we shall
summarize what we know about $\chi^*$.

First of all, because $\chi^*$ is unitary with respect to the
Liouville $L^2$--measure on phase space, the spectrum of $\chi^*$ lies
on the unit circle in ${\Bbb C}$.  (If the map is mixing, the spectrum
is known to be absolutely continuous \cite{arnold}.)  Second, a
salient feature of this spectrum is its {\it instability} with respect
to regularization by smoothing.  More precisely, if we project
${\chi^*}$ from the full $L^2$-space to a subspace of smooth
functions, by coarse graining the phase space or imposing an
ultraviolet (UV) cutoff, unitarity is lost and the spectrum moves to
the interior of the unit circle.  Third, if the map $\chi$ is mixing,
the spectrum of the projected operator consists of one isolated {\it
  nondegenerate} eigenvalue at $+1$ (with the corresponding
eigenfunction being the invariant density $\rho_0 = {\rm const}$) and
so--called Ruelle resonances \cite{ruelle} inside the unit circle.  It
is important that this feature persists, i.e.  the unitary spectrum is
{\it not} recovered, when the UV cutoff is lowered to zero.  (This is
a statement of non--interchangability of limits.)  Note also the
difference between the quantum and classical cases: the quantum
operator $1 - {\rm Ad}(U_N)$ always has $N$ zero modes, whereas its
classical limit $1 - \chi^*$ on smooth functions only has a {\it
  single} zero mode for a mixing system, namely the invariant density.

The usual approach taken in the study of Ruelle resonances is to
regularize in the ultraviolet by imposing a smoothness or
differentiability condition on the phase space functions acted upon by
the Frobenius--Perron operator.  A similar effect is achieved if we
leave the functions unchanged and, instead, implement the UV
regularization on the Frobenius--Perron operator {\it itself}.  For
our purposes the latter procedure is more convenient.  What we do is
to make the replacement
       \[ 
        \chi^* \; \longrightarrow \; r^\epsilon \chi^* \;, 
        \]
where the operator $r^\epsilon$ is chosen to be the exponential of a
sum of squares of Hamiltonian vector fields $\Xi_k$, 
        \[ 
        r^\epsilon = \exp \; \epsilon \sum_{k=1}^s \Xi_k^2 \;.  
        \]
For example, for $M = {\rm S}^2$ we take for $\Xi_k$ the generators of
rotations $(k = 1,2,3)$, in which case $\sum_{k=1}^3 \Xi_k^2$ is
simply the Laplacian on ${\rm S}^2$.  More generally, the choice of
Hamiltonian vector fields is constrained by the requirement that
$\sum_k \Xi_k^2$ be an elliptic operator.  The effect of $r^\epsilon$,
then, is to regularize in the ultraviolet by suppressing fluctuations
on short scales.  It is important that $r^\epsilon$ is the exponential
of a differential operator of {\it second} order, whereas the
Frobenius--Perron operator can be thought of as the exponential of a
first--order differential operator.  Therefore, $r^\epsilon$ is always
relevant in the ultraviolet, no matter how small is the value of
$\epsilon$.  As a result, $r^\epsilon$ remains effective as a
regulator even in the limit $\epsilon\to 0$ and, again, the unitary
spectrum is not retrieved when the UV cutoff is removed.

In the following we will make use of the fact that, if the map $\chi$
is mixing, the operator $1 - r^\epsilon \chi^*$ is strictly positive
on the orthogonal complement of the uniform state $\rho_0$.  The
physical reason for such behavior is this.  Consider the long time
dynamics obtained by iterating $r^\epsilon \chi^*$ and, for simplicity
of the argument, assume $\chi$ to be uniformly hyperbolic, which means
that $\chi$ is everywhere stretching along unstable manifolds and
contracting along stable ones.  Thanks to the ellipticity of
$\sum_{k=1}^r \Xi_k^2$, the operator $r^\epsilon$ counteracts the
contraction along the stable manifolds via diffusion.  After many
iterations, the combination of stretching by $\chi$ and smoothing by
$r^\epsilon$ will attract any initial state to the uniform state.
Thus, as time $l$ goes to infinity, the operator $(r^\epsilon
\chi^*)^l$ approaches the projector onto the uniform state $\rho_0$.
This implies strict positivity of the real part of the spectrum of $1
- r^\epsilon \chi^*$ operating on the orthogonal complement of
$\rho_0$.

Let us now turn from the space of functions on $M$, acted upon by
$\chi^*$, to the quantum mechanical space ${\cal H}_N \otimes {\cal
  H}_N^*$, acted upon by the operator ${\rm Ad}(U_N)$.  Crudely
speaking, the quantum space differs from the classical one by a finite
resolution, or cutoff, set by Planck's constant.  When this cutoff is
much smaller than the regularization scale set by $\epsilon$, we have
a good semiclassical expansion, (\ref{moyal}), and the regularized
${\rm Ad}(U_N)$ spectrum looks qualitatively similar to the
regularized Frobenius--Perron spectrum.  On the other hand, if $\hbar$
is kept fixed at a finite value, the destruction of the unitary
spectrum of ${\rm Ad}(U_N)$ with increasing $\epsilon$ is not abrupt
but is rounded off.  If we take $\epsilon \sim \hbar^\alpha$ with
$\alpha > 2d$, the strength of the perturbation caused by the
regulator becomes smaller than the level spacing of ${\rm Ad}(U_N)$
[which is $2\pi/N^2 \sim \hbar^{2d}$], and is therefore negligible, in
the limit $\hbar\to 0$.  In contrast, if we choose the scaling
exponent $\alpha$ in the range $0 < \alpha < 1$, then by power
counting the regulator remains relevant for $\hbar\to 0$.  This is
good news: it suggests that we can kill the unitary spectrum of ${\rm
  Ad}(U_N)$, and thereby get rid of the zero mode problem if the
classical map is mixing, by a regulator with strength $\epsilon$ that
{\it vanishes} in the classical limit.

The basic strategy to follow should now be clear: our aim must be to
substitute for the unitary operator ${\rm Ad}(U_N)$ in the quadratic
functional $S^{(2)}$ the regularized operator ${R}_N^\epsilon {\rm
  Ad}(U_N)$, where $R_N^\epsilon$ is a quantization of the diffusion
operator $r^\epsilon$.  Such a substitution is expected to turn the
unitary spectrum of ${\rm Ad}(U_N)$ into something akin to the
Frobenius--Perron spectrum of the classical operator ${\chi^*}$.  If
$\chi$ is mixing and $\epsilon$ depends on $\hbar$ as $\epsilon \sim
\hbar^\alpha$ with $0 < \alpha < 1$, only the invariant zero mode of
$1-{\rm Ad}(U_N)$ remains neutral.  The remaining $N-1$ zero modes
turn into positive directions of the regularized Hessian and therefore
become amenable to perturbative treatment.  From experience with the
application of the supersymmetric method to disordered electron
systems \cite{efetov}, we then anticipate that a careful integration
over the invariant zero mode will give the random matrix answer for
the correlation functions.

What is unsatisfactory about this strategy is the absence of any
explanation of {\it where the UV regularization came from}. So far,
regularization seems to be an ad hoc procedure, or ``dirty trick'',
introduced so as to yield the result one wants to get.  The situation
would be different if we were working in a quantum field theoretic
context.  In that case, UV regularization would be well justified by
the fundamental and inevitable existence of ultraviolet singularities
(signaling incompleteness of the field theory, or ``new physics'' at
short scales), which must be tamed by the introduction of a cutoff.
In contrast, our formula (\ref{susy_formula}) has no singularities
whatsoever but is perfectly well--defined -- we just don't know how to
evaluate it -- and there is no a priori need for regularization.  In
passing we mention that, if we ignored the arguments of
Sect.~\ref{sec:semiclassics} forbidding us to apply the semiclassical
expansion (\ref{moyal}) to the action functional $S$, and applied the
expansion anyway, then we would, in fact, end up with an ultraviolet
singular field theory \cite{a3s}.  However, the UV divergencies so
introduced are entirely artificial -- they simply confirm the fact
that we have made improper use of the semiclassical expansion -- and
do not justify the postulate of any regulator.

To reiterate, it is illegal just to postulate UV regularization as a
technical device or trick.  For a convincing argument, one needs to
explain exactly what modification of the original problem is implied
by regularization, and one has to do that in a frame that precedes the
supersymmetric quantum field theory formalism.  As will be argued in
the remainder of this section, regularization of the Hessian of
(\ref{action_S}) at $Z = \tilde Z = 0$ amounts to averaging over the
``quantum'' ensemble described in Sect.~\ref{sec:conjecture}.

Recall that, if $\xi_k$ $(k=1,...,s)$ are uncorrelated Gaussian random
variables with zero mean and variance $\langle \xi_k^2 \rangle =
2 \epsilon$, the ensemble is specified by
           \[
           U_N(\xi) = \exp \left( i \sum_{k=1}^s
          \xi_k X_{k,N} / \hbar \right) U_N \;.
           \]
By averaging the Hessian $1 - {\rm Ad} \left( U_N(\xi) \right)$ over 
the probability space of the parameters $\xi_k$, we get
       \[
       \left\langle 1 - {\rm Ad} \left( U_N(\xi) \right) \right\rangle
       = 1 - {R}_N^\epsilon {\rm Ad}(U_N) \;,
       \]
where the operator $R_N^\epsilon$ is identified as
       \begin{eqnarray}
       {R}_N^\epsilon &=& \left\langle 
        {\rm Ad} \left( \exp \; i \sum_{k=1}^s \xi_k X_{k,N} / \hbar \right)
        \right\rangle 
        \nonumber \\
        &=& \left\langle
        \exp \; i \sum_k \xi_k {\rm ad}(X_{k,N}) / \hbar 
        \right\rangle 
        = \exp \; - \epsilon \sum_k {\rm ad}^2(X_{k,N}) / \hbar^2 
        + ... \;. \nonumber
       \end{eqnarray}
[Note that, since the integrand in (\ref{susy_formula}) depends on
$U_N$ in a nonlinear way, later we will have to confront averages of
powers of ${\rm Ad} \left( U_N(\xi) \right)$.  For the time being, we
concentrate on the Hessian.]  The terms indicated by dots in the last
expression become negligible in the semiclassical limit $\hbar \to 0$.
Taking the commutator with $X_{k,N}$ in this limit corresponds to 
applying the Hamiltonian vector field $\Xi_k$:
       \[
       {\rm ad}(X_{k,N}) = [ X_{k,N} , \bullet ] \;
       {\buildrel {\hbar\to 0} \over \longrightarrow} \;
       i\hbar \Xi_k \;,
       \]
so $R_N^\epsilon$ is in fact a quantization of the classical diffusion
operator $e^{\epsilon \sum_{k=1}^s \Xi_k^2} = r^\epsilon$, as
desired.

To conclude, averaging over the specified ensemble regularizes the
Hessian in the ultraviolet.  As was argued earlier, if the classical
map $\chi$ is mixing the real part of $1 - r^\epsilon\chi^*$ is
positive on all states that are orthogonal to the uniform state
$\rho_0$.  In view of the semiclassical limits $R_N^\epsilon \to
r^\epsilon$ and ${\rm Ad}(U_N) \to \chi^*$, we expect the same to be
true for the regularized Hessian $1 - R_N^\epsilon {\rm Ad} \left( U_N
\right)$ on ${\cal D}_N^\perp$, the subspace of ${\cal H}_N \otimes
{\cal H}_N^*$ orthogonal to the invariant element.  This key
observation allows us to proceed as follows.

\section{Asymptotic expansion}
\label{sec:asymptotics}

We set $\gamma_\pm = e^{\pm ic_\pm / N}$ where $c_\pm = {\rm diag}
(c_{\pm 0},c_{\pm 1})$ are $2 \times 2$ matrices and, to abbreviate
the notation, we put $Z_U := (1_2 \otimes U_N) Z (1_2 \otimes
U_N^{-1})$.  Then, since $c_\pm$ are held fixed while $N$ is sent to
infinity, we may expand the action functional:
      \begin{eqnarray}
        S &=& - {\rm STr} \ln (1 - \tilde Z Z) 
        + {\rm STr} \ln ( 1 - \tilde Z Z_U )
        \nonumber \\
        &&- {i\over N} {\rm STr} \left( 
        c_+ Z_U \tilde Z (1 - Z_U \tilde Z)^{-1} - 
        c_- \tilde Z Z_U (1 - \tilde Z Z_U)^{-1} \right) + {\cal O}(1/N^2)
        \;. \nonumber
      \end{eqnarray}
To isolate the invariant zero mode, we make the substitution
      \begin{eqnarray}
        Z &=& (\zeta + Z_0)(1 + \tilde Z_0 \zeta)^{-1} \;,
        \nonumber \\
        \tilde Z &=& (\tilde\zeta + \tilde Z_0)(1 + Z_0 \tilde\zeta)^{-1} \;.
        \nonumber
      \end{eqnarray}
The supermatrices $\zeta$ and $\tilde\zeta$ are taken to be traceless 
in each block (i.e. ${\rm Tr}\; \zeta^{\sigma\tau} = {\rm Tr}\; \tilde
\zeta^{\sigma\tau} = 0$ for $\sigma , \tau = {\rm B}, {\rm F})$, and
the zero mode is parameterized by $Z_0 = z_0 \otimes 1_N$, $\tilde Z_0
= \tilde z_0 \otimes 1_N$ with $2 \times 2$ matrices
       \[
        z_0 = \pmatrix{z_0^{\rm BB} &z_0^{\rm BF}\cr z_0^{\rm FB}
        &z_0^{\rm FF}\cr} \;, \quad
        \tilde z_0 = \pmatrix{\tilde z_0^{\rm BB} &\tilde z_0^{\rm BF}\cr 
        \tilde z_0^{\rm FB} &\tilde z_0^{\rm FF}\cr} \;.
       \]
Because the transformation
       \begin{eqnarray}
        Z &\mapsto& (AZ+B)(CZ+D)^{-1} \;,
        \nonumber \\
        \tilde Z &\mapsto& (C+D\tilde Z)(A+B\tilde Z)^{-1} \;,
        \nonumber
       \end{eqnarray}
is an isometry of the Riemannian symmetric superspace, the Berezinian
of the change of variables from $Z$ to $(\zeta,Z_0)$ is unity and
$D(Z,\tilde Z)$ factors into a product of Berezin forms $D(z_0,\tilde
z_0) D(\zeta,\tilde\zeta)$.  (We pay no attention to possible boundary
anomalies here.) Now, by inserting the expansion of the action $S$
into (\ref{susy_formula}), changing variables as indicated, and using 
identities such as 
       \[
        (1-Z_U\tilde Z)^{-1} = (1+Z_0\tilde\zeta)
        (1-\zeta_U\tilde\zeta)^{-1} (1+\zeta_U\tilde Z_0)
        (1 - Z_0 \tilde Z_0)^{-1} \;,
       \]
we obtain
       \begin{eqnarray}
        \Omega(e^{ic_+/N};e^{-ic_-/N}) &=& 
        \int D(z_0,\tilde z_0) \; \Omega_1(c_+/N,c_-/N;z_0,\tilde z_0)
        \nonumber \\
        &&{\hspace{-1cm}}\times \exp \; i {\rm STr}_{{\Bbb C}^{1|1}} 
        \left( c_+ z_0 \tilde z_0 (1 - z_0 \tilde z_0)^{-1} - 
        c_- \tilde z_0 z_0 (1 - \tilde z_0 z_0)^{-1} \right)
        \nonumber
       \end{eqnarray}
with $\Omega_1$ defined by
       \begin{eqnarray}
        &&\Omega_1(c_+/N,c_-/N;z_0,\tilde z_0) 
        \nonumber \\
        &=& \int D(\zeta,\tilde\zeta)
        \; {\rm SDet}(1-\tilde\zeta\zeta) 
        {\rm SDet}^{-1}(1-\tilde\zeta\zeta_U) 
        \nonumber \\
        &&\times \exp \Big( {i\over N} 
        {\rm STr} \; c_+ (1+Z_0\tilde\zeta)
        (1-\zeta_U\tilde\zeta)^{-1}\zeta_U(\tilde\zeta+\tilde Z_0)
        (1-Z_0\tilde Z_0)^{-1}
        \nonumber \\
        &&{\vphantom{ \exp \Big( }} - {i\over N} {\rm STr} \; 
        c_- (1+\tilde Z_0\zeta_U) (1-\tilde\zeta\zeta_U)^{-1}
        \tilde\zeta(\zeta_U+Z_0) (1-\tilde Z_0 Z_0)^{-1} \Big) \;.
        \nonumber
       \end{eqnarray}
Note that convergence of the last integral is ensured by the condition
${\rm Im} c_{+0} > 0 > {\rm Im} c_{-0}$ resulting from $|\gamma_{\pm
  0}| < 1$.  

Our next goal is to establish an asymptotic $1/N$ expansion for the
function $\Omega_1$.  To that end, we temporarily set $c_+/N = c_-/N =
0$ and consider the integral
      \[
        \Omega_1(0,0;z_0,\tilde z_0) = \int D(\zeta,\tilde\zeta) \;
        {\rm SDet}(1-\tilde\zeta\zeta) {\rm SDet}^{-1} \left( 1 - 
        \tilde\zeta \zeta_U \right) \;.
      \]
To compute this, we observe that both the Berezin form $D(\zeta,
\tilde\zeta)$ and the two superdeterminants are invariant under
transformations
      \[
        \zeta \mapsto A \zeta D^{-1}, \quad
        \tilde\zeta \mapsto D \tilde\zeta A^{-1}
      \]
with $A = a \otimes 1_N$ and $D = d \otimes 1_N$.  (Indeed, such
transformations commute with ${\rm Ad}(U_N)$ and are isometries of the
symmetric superspace.)  This invariance results in the value of the
integral being equal to the value of the integrand at the origin
$\zeta = \tilde\zeta = 0$:
       \[
        \Omega_1(0,0;z_0,\tilde z_0) = 1 \;,
       \]
by what is often called the Parisi--Sourlas--Efetov--Wegner theorem in
disordered electron physics.  The mechanism underlying the theorem is
called ``localization'' in the mathematics literature.  In a setting
closely related to equivariant cohomology, localization of
superintegrals has recently been discussed by Schwartz and Zaboronsky
\cite{sz}.  With reference to their results we can argue as follows.
On setting
       \[
       A = \exp \pmatrix{0 &\alpha\cr \bar\alpha &0\cr} \otimes 1_N \;,
       \qquad
       D = \exp \pmatrix{0 &\delta\cr \bar\delta &0\cr} \otimes 1_N \;,
       \]
and differentiating with respect to the Grassmann parameters $\alpha,
\bar\alpha, \delta, \bar\delta$, the transformation $\zeta \mapsto
A\zeta D^{-1}$, $\tilde\zeta \mapsto D \tilde\zeta A^{-1}$ gives rise
to four odd vector fields, each of which leaves the integrand
invariant.  The joint zero locus of these vector fields is $\zeta =
\bar\zeta = 0$, and the superdeterminant of the Hessian at this point
is unity by cancellation due to supersymmetry.  Moreover, the vector
fields are {\it compact} in the sense of Schwartz and Zaboronsky.
Therefore, by the version of the Parisi--Sourlas--Efetov--Wegner
theorem proved in \cite{sz}, the integral equals unity as claimed.

Given the result $\Omega_1(0,0;z_0,\tilde z_0) = 1$, one would like to
proceed and make a saddle point approximation around $\zeta = \tilde
\zeta = 0$, to produce an asymptotic expansion for $\Omega_1 (c_+/N,
c_-/N; z_0, \tilde z_0)$ of the form $1 + N^{-1} f(c_+,c_-;z_0,\tilde
z_0) + {\cal O}(N^{-2})$.  However, this is impossible as it stands.
The reason is the notorious problem of $N-1$ zero modes: there exists
an ($N-1$)--dimensional maximal torus of group actions that commute
with ${\rm Ad}(U_N)$ and leave ${\rm SDet}(1-\tilde\zeta\zeta){\rm
  SDet}^{-1}(1-\tilde\zeta \zeta_U)$ invariant, thereby causing a lack
of stability of the candidate saddle point $\zeta = \tilde\zeta = 0$.
Fortunately, given all the preparations that were made in
Sect.~\ref{sec:regularization}, we immediately know how to fix this
stability problem.

We substitute $U_N(\xi)$ for $U_N$ and average over the probability
space of the ensemble, replacing $\Omega$ by $\langle\Omega\rangle$,
and $\Omega_1$ by $\langle\Omega_1\rangle$.  The integral for
$\Omega_1$ converges and, as a matter of fact, converges {\it
  uniformly} in the parameters $\xi_k$, if ${\rm Im} c_{+0} \ge 0 \ge
{\rm Im} c_{-0}$.  Therefore, we may interchange the order of
integration and ensemble averaging.  By the same localization 
argument as before,
       \[
        \left\langle \Omega_1(0,0;z_0,\tilde z_0) \right\rangle = 
        \int D(\zeta,\tilde\zeta) \left\langle{\rm SDet}(1-\tilde\zeta\zeta)
        {\rm SDet}^{-1}(1-\tilde\zeta \zeta_{U_N(\xi)}) \right\rangle
        = 1 \;.
       \]
What has improved in comparison with the situation without ensemble 
averaging is that now we do have a good asymptotic expansion.  The
reason is that integrals such as
       \begin{eqnarray}
        &&\int D(\zeta,\tilde\zeta) \;
        \Big\langle {\rm SDet}(1-\tilde\zeta\zeta) 
        {\rm SDet}^{-1}(1-\tilde\zeta\zeta_{U_N(\xi)}) 
        \nonumber \\
        &&\times {\rm STr} \; c_+ 
        (1-\zeta_{U_N(\xi)}\tilde\zeta)^{-1}
        \zeta_{U_N(\xi)} \tilde\zeta (1-Z_0\tilde Z_0)^{-1}
        \Big\rangle
        \nonumber
       \end{eqnarray}
are expected to exist and remain finite in the limit $N \to \infty$.
This can be checked by expanding around the saddle point $\zeta =
\tilde\zeta = 0$ in the usual manner.  By doing the resulting Gaussian
integrals over $\zeta,\tilde\zeta$, one obtains traces involving
powers of the inverse of the regularized Hessian $1 - R_N^\epsilon
{\rm Ad}(U_N)$ acting on ${\cal D}_N^\perp$.  We have argued earlier
that the regularized Hessian remains strictly positive in the
semiclassical limit if the classical map is mixing.  Therefore, the
inverse of $1 - R_N^\epsilon {\rm Ad}(U_N)$ exists on ${\cal
  D}_N^\perp$ and is uniformly bounded in $N$.  Moreover, as follows
from semiclassical trace formulas, the traces encountered are bounded
uniformly in $N$ if the fixed points of the classical map are
isolated.

What all this amounts to is that, unlike $\Omega_1$, the ensemble 
average $\langle \Omega_1 \rangle$ does have an asymptotic expansion:
       \[
       \left\langle \Omega_1(c_+/N,c_-/N;z_0,\tilde z_0) \right\rangle 
       = 1 + N^{-1} f(c_+,c_-;z_0,\tilde z_0) + {\cal O}(N^{-2}) \;.
       \]
Using this in the formula for the correlator $\langle\Omega\rangle$
we obtain
       \begin{eqnarray}
       &&\lim_{N\to\infty} \left\langle \Omega(e^{ic_+/N};e^{-ic_-/N})
         \right\rangle \nonumber \\
         &=& \int D(z_0,\tilde z_0) \exp \; i 
         {\rm STr} \left( c_+ z_0 \tilde z_0 (1 - z_0 \tilde z_0)^{-1} - 
        c_- \tilde z_0 z_0 (1 - \tilde z_0 z_0)^{-1} \right) \;.
       \nonumber
       \end{eqnarray}
Computation of this definite superintegral is a standard exercise
and gives the random matrix answer (\ref{rmt}).  The result for the
pair correlation function follows by taking two derivatives.  The
dependence on $z_0, \tilde z_0$ of the $1/N$ correction to $\langle
\Omega_1 \rangle$ can be shown to be such that the integral over $z_0,
\tilde z_0$ exists and the coefficient of $1/N$ in $\langle \Omega
\rangle$ is finite.

In conclusion, I believe that the arguments presented here are strong
evidence in favor of the precise version of the BGS conjecture
proposed.  To turn them into a proof that is rigorous by mathematical
standards, three improvements are required.
\begin{enumerate}
\item One needs to do a spectral analysis of the operator
  $R_N^\epsilon {\rm Ad}(U_N)$ to support our intuition of what
  happens in the limit $N \to \infty$.
\item The transformation of Berezin forms $D(Z,\tilde Z) \to
  D(z_0,\tilde z_0) D(\zeta,\tilde\zeta)$ needs to be scrutinized
  carefully, to rule out the possible existence of boundary anomalies
  (Sect.~\ref{sec:basics}) that might interfere with our argument.
\item A complete computation of the $1/N$ corrections is called for.
In particular, one must resolve the puzzle why the calculation of
Andreev et al, which is closely related to ours, fails to reproduce
the diagonal approximation to the double sum over periodic orbits.
\end{enumerate}

{\bf Acknowledgment}. While participating in the workshop on
``Disordered Systems and Quantum Chaos'' at the Newton Institute,
Cambridge, where these lectures were delivered, I enjoyed a number of
illuminating discussions with Steve Zelditch, who got me straightened
out on the subject of semiclassical limits and suggested to compose
the deterministic map with a stochastic Hamiltonian flow that averages
to a diffusion operator.

\end{document}